\documentclass[twocolumn,amssymb,prb]{revtex4}
\usepackage{epsfig}% Include figure files
\usepackage{graphicx}% Include figure files

\begin{document}

%\begin{frontmatter}

\title{Quantum dot detects Majorana modes of both chiralities}

\author{Piotr Stefa\'nski}
\address{Institute of Molecular Physics of the Polish Academy of Sciences\\
  ul. Smoluchowskiego 17, 60-179 Pozna\'n, Poland}

\begin{abstract}
A tunneling junction between normal electrode and a topological superconducting wire, mediated by a quantum dot, is considered theoretically.  We show that the presence of the dot in the junction can be advantageous to Majorana zero modes identification. Namely, we demonstrate that for the dot strongly coupled to the wire, the Majorana mode from the upper chiral sub-band "leaks" into the dot, providing supplementary information on Majorana mode formation. Thus, both the Zeeman-split dot sub-levels detect Majorana partners of a Kramers pair, formed at the wire end. The characteristic three-peak structures in both spin sectors of the spectral density of the dot, distinguish from the trivial scenario of one Andreev resonance at Fermi energy produced exclusively by the dot's spin sub-levels.
\end{abstract}
\maketitle
%\end{frontmatter}

\section{Introduction}
Majorana fermions \cite{EttoreMajorana1937} recently attracted attention  due to a promising potential possibility of coherence undistorted quantum computation \cite{DasSarma2015}. The early theoretical proposals \cite{Lutchyn2010b,Oreg2010a} induced enormous experimental efforts to realize Majorana quasiparticles in a solid state device \cite{Lutchyn2018}. The most suitable for "braiding" operations on Majorana zero modes (MZMs) appears to be their implementation in quasi-$1D$ superconductor(Al, NbTiN)-semiconductor(InAs, InSb) heterostructure with strong spin-orbit interaction, subjected to external magnetic field  \cite{Marra2022d}. However, recent experimental findings \cite{Valentini2021} cast a serious doubt on previous tunneling spectroscopy measurements. Namely, it was shown that an  unintentionally formed quantum dot (QD) inside the junction can introduce misleading, similar to MZM, trivial Andreev bound states close to Fermi energy. Thus, zero bias peaks (ZBPs) in differential conductance are not sufficient to prejudge on MZMs, additional complementary evidences are needed. The most convincing is the evidence of closing and reopening of the superconducting gap in the wire, as an indication of the transition to the topological superconducting phase. A suitable screening protocol has been recently proposed \cite{Pikulin2021} for the three-terminal device configuration with two normal leads and one superconducting lead. This protocol has been successfully experimentally fulfilled lately in InAs-Al devices \cite{Aghaee2022}. However, it has been demonstrated theoretically  \cite{Hess2023} that such superconducting gap reopening can be mimicked by trivial Andreev sub-bands if the length of the wire is not significantly larger than superconducting coherence length. Thus, the efforts in seeking of reliable supplementary evidences on MZMs formation are continued.

In this work we propose to take into account another supplementary feature which can help to distinguish true Majorana zero modes from the artificial ones. We consider a setup similar to the experimental one \cite{Valentini2021}, and demonstrate that the formation of unintentional quantum dot in the tunneling junction can be advantageous in MZM detection. Indeed, a quantum dot formed in the junction has been also experimentally utilized to probe sub-gap states in the wire \cite{Deng2016a} and to estimate the degree of nonlocality of Majorana modes \cite{Deng2018}.

In discussion of our results we exploit the fact that in the topological wire  two chiral sub-bands are formed, each of them introducing MZM at the end of the wire. This MZM pair originates from the Majorana Kramers pair forming in time-reversal invariant superconductors \cite{Zhang2013}. The formation of chiral sub-bands is an evidence of strong spin-orbit Rashba interaction in the semiconductor-superconductor wire heterostructure, which otherwise can be substantially diminished \cite{Reeg2018a}. Strong Rashba field is crucial for possible Majorana modes existence. Experimentally, the detection of chiral sub-bands provides a supplementary information, which in conjunction with observation of zero-bias structures in differential conductance for both QD spin sub-levels, when tuned to Fermi energy, yields a stronger evidence for true MZMs.

Contrary to the typical, widely considered scenario of a single chiral sub-band coupled to a given QD's spin sub-level \cite{Liu2011a,Vernek2014b,Ruiz-Tijerina2015,Liu2017c,Ptok2017,Prada2017a,Chevallier2018a,Bittermann2022,Liu2022b,Diniz2023}, we demonstrate that under certain conditions, characteristic to unintentional QD formation, the MZM from the second chiral sub-band can also be detected.

\section{The model and calculation approach}
Quantum dot is described by the Hamiltonian:
\begin{equation}\label{H_QD}
H_{QD}=\sum_{\sigma=\downarrow,\uparrow}\epsilon_{d\sigma}d_{\sigma}^{\dagger}d_{\sigma},
\end{equation}
where the spin sub-levels are split by external magnetic field: $\epsilon_{d\downarrow/\uparrow}=\epsilon_{d}\mp V_{z}^{QD}$ and can be shifted by gate voltage.
The dot is assumed to be strongly coupled to the wire, with not a distinctly defined confinement; a probable scenario for its formation in the junction. Such large dots have their spatial levels one close to the other in an energy scale. When an external magnetic field is applied, the Zeeman split spin sub-levels $\epsilon_{\downarrow}$ and $\epsilon_{\uparrow}$ from two adjacent spatial levels can be close together and simultaneously enter the superconducting gap. Another possible scenario is for a single spin-split QD's level with reduced splitting due to a  $g$-factor different from the one in the wire. Its value depends on the interplay between Rashba and Dresselhaus spin-orbit interactions, applied gate voltage and the dot radius \cite{DeSousa2003}. Due to partially delocalized nature of the dot, it is assumed that the on-site Coulomb interactions can be neglected and the dot is effectively non-interacting.

In order to perform tunneling spectroscopy, the dot is coupled to the normal electrode, which is described by the Hamiltonian:
\begin{equation}
H_{lead}+H_{tun}=\sum_{k,\sigma=\downarrow,\uparrow}c_{k\sigma}^{\dagger}c_{k,\sigma}+\sum_{k,\sigma}\left[t_{l}c_{k\sigma}^{\dagger}d_{\sigma}+h.c.\right],
\end{equation}
and introduces a QD level broadening $\Gamma_{l}$.
Finally, the dot is coupled to the first site of the wire:
\begin{equation}
H_{QD-W}=\sum_{\sigma=\downarrow,\uparrow}\left[t_{w} d_{\sigma}^{\dagger}c_{1\sigma}+h.c.\right]
\end{equation}

The semiconducting wire with large spin-orbit interaction and induced $s$-wave superconductivity is modelled by the tight-binding Hamiltonian (see \cite{Stefanski2021}) with no on-site Coulomb interactions. Additionally, the wire is subjected to an external magnetic field, which when exceeding the critical value $V_{z}^{cr}=\sqrt{\mu^{2}+\Delta^{2}}$, can drive the wire into the topological state. In order to obtain the maximal effect in MZM formation in the wire and the well resolved chiral sub-bands, the external magnetic field and effective Rashba field in the wire Hamiltonian are assumed to be perpendicular \cite{Mourik2012}.

We calculate the Dyson equation matrix for the retarded Green's function of the dot:

\begin{equation}\label{GdMat}
\hat{G}_{d}=\left[\hat{g}_{0}^{-1}-\hat{\Sigma}_{l}-\hat{\Sigma}_{wire} \right]^{-1},
\end{equation}
where $\hat{g}_{0}$ is the bare QD Green's function matrix written in Nambu basis $\Psi=(d_{\downarrow},d_{\uparrow},d_{\uparrow}^{\dagger},d_{\downarrow}^{\dagger})$. It is diagonal with the elements: $\hat{g}_{0}[1,1]=(\omega-\epsilon_{d\downarrow})^{-1}$, $\hat{g}_{0}[2,2]=(\omega-\epsilon_{d\uparrow})^{-1}$, $\hat{g}_{0}[3,3]=(\omega+\epsilon_{d\uparrow})^{-1}$ and $\hat{g}_{0}[4,4]=(\omega+\epsilon_{d\downarrow})^{-1}$.

The self-energy matrices $\hat{\Sigma}_{l}$ and $\hat{\Sigma}_{wire}$ describe the coupling to the normal lead and  to the wire, respectively. The $\hat{\Sigma}_{l}$ matrix is diagonal with the elements equal to $-i\Gamma_{l}$. The coupling to the wire is described by the self-energy:
\begin{equation}\label{Sig_wire}
\hat{\Sigma}_{wire}=\hat{t}_{w}[\hat{G}]_{1,1}\hat{t}_{w}^{*},
\end{equation}
where $\hat{t}_{w}$ matrix describes the coupling of the dot to the first site of the wire. It is diagonal with elements: $\hat{t}_{w}[1,1]=\hat{t}_{w}[2,2]=-\hat{t}_{w}[3,3]=-\hat{t}_{w}[4,4]=t_{w}$.

The Green's function matrix $[\hat{G}]_{1,1}$ represents the first site of the wire, and is calculated by recursive summation over the full length of the wire  \cite{Stefanski2021}.

In the following we calculate the spectral density of the dot: $\rho_{QD}=-(1/\pi)Im[Tr[\hat{G}_{d}]]$, which can be measured experimentally  by tunneling spectroscopy.

\section{Numerical results}
The Majorana zero mode produces a quantized peak of the height $2e^{2}/h$  in zero bias conductance \cite{Flensberg2010b}, which at zero temperature  is robust against variations of the
tunneling strength between the quantum dot and the wire, the positions of the dot level and, the wire chemical potential. In our case the quantized peak (multiplied by $\pi\Gamma_{l}$) is slightly below the predicted value due to added imaginary dissipation term $\delta=0.005$ in the wire for the smoothing its density of states.

\begin{figure} 
\centering
\includegraphics[width=9cm]{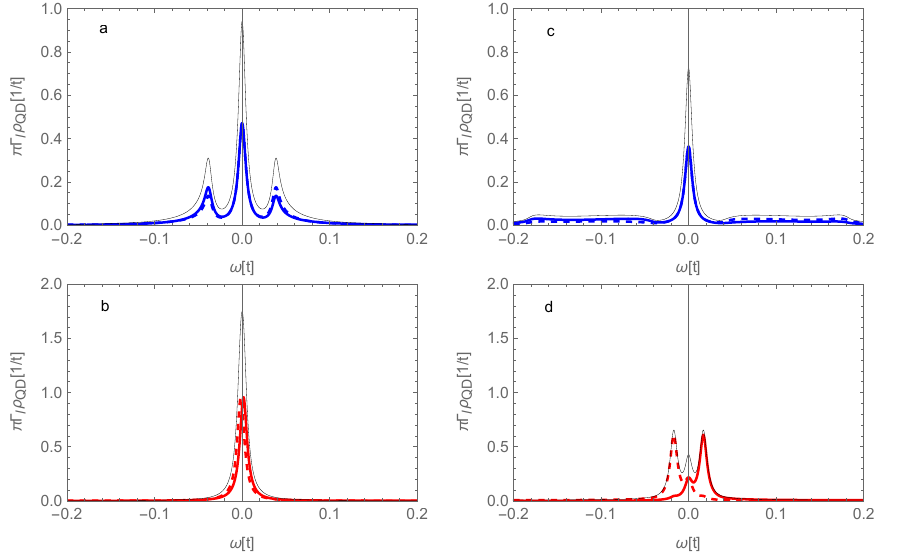}
\caption{\label{fig1U=0}Spectral densities of the dot. Solid (dashed) line - particle (hole) parts, thin solid line - total spectral densities. Panels (a) and (c) are for spin-$\downarrow$ sector when the dot spin level $\epsilon_{d\downarrow}=\epsilon_{F}$, while panels (b) and (d) are for spin-$\uparrow$ sector when the dot spin level $\epsilon_{d\uparrow}=\epsilon_{F}$. Panels (a) and (b) are for the dot weakly coupled to the wire for $t_{w}=0.20$. Panels (c) and (d) are for the dot strongly coupled to the wire for $t_{w}=0.44$. Remaining input parameters are: $t=1$, $\mu=0$, $\Delta=0.2$, $t_{so}=2\Delta$,  $V_{z}=1.2V_{z}^{cr}$, $V_{z}^{QD}=0.5 V_{z}$, $\Gamma_{l}=0.005$ and $\delta=0.005$.}
\end{figure}

In Fig.~(\ref{fig1U=0}) we present numerical results for QD's spectral density. The magnetic field in the wire  is set to $V_{z}=1.2 V_{z}^{cr}$, and it induces the topological phase. The spin sub-levels of the dot are subsequently tuned by an appropriate gate voltage to Fermi level.
 Spectral densities of the dot for  $\epsilon_{\downarrow}=\epsilon_{F}$ are shown in the upper row (panels (a) and (c)),
and for $\epsilon_{\uparrow}=\epsilon_{F}$ in the lower row (panels (b) and (d)). For $\epsilon_{\downarrow}=\epsilon_{F}$ ( $\epsilon_{\uparrow}=\epsilon_{F}$) the spectral density is dominated by its spin-down (spin-up) components: $\rho_{QD}\simeq -(1/\pi)Im[\hat{G}_{d}[1,1]+\hat{G}_{d}[4,4]]=\rho_{\downarrow}^{p}+\rho_{\downarrow}^{h}$ ( $\rho_{QD}\simeq -(1/\pi)Im[\hat{G}_{d}[2,2]+\hat{G}_{d}[3,3]]=\rho_{\uparrow}^{p}+\rho_{\uparrow}^{h}$), where $p$ and $h$ stand for particle and hole components, respectively.

 For weak coupling of the dot to the wire (the results displayed in the left column)
 a characteristic three-peak structure is observed in the spin-down sector (panel (a)), with MZM accompanied by the satellite
peaks due to the split QD $\epsilon_{\downarrow}$, localized at $\omega\sim\mp(\epsilon_{\downarrow}+Re\Sigma_{wire}[1,1])$. Contrary, when $\epsilon_{\uparrow}$ is tuned to Fermi energy (panel (b)), only trivial Andreev bound state resonance at $\epsilon_{F}$ is observed.
Direction of the magnetic field applied to the system determines the lower, active chirality sub-band of the wire (see Section \ref{SecChiral}); for $V_{z}>0$ the MZM from the lower $"-"$-chiral sub-band
 "leaks" into the spin-down sector of the dot. The upper $"+"$-chiral sub-band
 is practically decoupled from the dot, which results in only trivial resonance in the spin-up spectral density of the dot for $\epsilon_{\uparrow}=\epsilon_{F}$ (panel (b)). This is the typically considered scenario of a single QD's spin sub-level coupled to an active chiral sub-band of the wire, equivalent to Kitaev chain \cite{Liu2011a}, described by the spinless Hamiltonian.
The picture changes for the dot strongly coupled to the wire (the results depicted in the right column). In the spin-down sector the Majorana resonance is still present,
but the satellite peaks are shifted apart, outside the gap, due to large $t_{w}$ hopping. In this case the MZM resonance can be hardly distinguished experimentally from a trivial Andreev resonance, which may develop at Fermi energy.  Now however, due to the large QD-wire coupling,
also the $"+"$-chiral sub-band becomes coupled to the dot, and the second member of the Majorana  pair adjacent to the dot can "leak" into spin-up sector of the dot;
 it is manifested by characteristic three-peak structure for $\epsilon_{\uparrow}=\epsilon_{F}$, panel (d), with split QD Andreev bound states localized at $\omega\sim\mp(\epsilon_{\uparrow}+Re\Sigma_{wire}[2,2])$.

\section{\label{SecChiral}Hamiltonian of the wire in the chiral basis}
For the interpretation of the numerical results it is instructive to rewrite the wire Hamiltonian in chiral basis \cite{Alicea2010}. First we transform the Hamiltonian of the wire (see Eq.~(9) in \cite{Stefanski2021}) into $k$-space; $H=H_{0}+H_{so}+H_{sc}$:
\begin{eqnarray}\label{H_in_k}
\nonumber
 H_{0}=\sum_{k,\sigma}\tilde{\epsilon}_{k\sigma}c^{\dagger}_{k\sigma}c_{k\sigma},\\
 H_{so}= \sum_{k}(\tilde{t}_{so}c^{\dagger}_{k\uparrow}c_{k\downarrow}+\tilde{t}_{so}^{\star}c^{\dagger}_{k\downarrow}c_{k\uparrow}),\\\nonumber
 H_{sc}=\sum_{k}(\Delta c_{-k\downarrow}c_{k\uparrow}+\Delta^{\star}c_{k\uparrow}^{\dagger}c_{-k\downarrow}^{\dagger}),
\end{eqnarray}
where $\tilde{\epsilon}_{k\sigma}=\bar{\mu}\mp V_{z}$ for spin $\sigma=\downarrow$ and $\sigma=\uparrow$, respectively, $\bar{\mu}=-\mu+4t\sin^{2}(k a/2)$ and $\tilde{t}_{so}=2 i t_{so}\sin(k a)$. In the next step we diagonalize the Hamiltonian $H_{0}+H_{so}$ in the spinor basis $\psi=(c_{k\uparrow}, c_{k\downarrow})$:
\begin{eqnarray}
H_{ch}=\left(
\begin{array}{cc}
\tilde{\epsilon}_{k\uparrow} & \tilde{t}_{so}\\
\tilde{t}_{so}^{\star} & \tilde{\epsilon}_{k\downarrow}
\end{array}
\right).
\end{eqnarray}
Diagonalization yields the chiral eigen-functions: $\epsilon_{k\mp}=\bar{\mu}\mp R$ with $R=\sqrt{|\tilde{t}_{so}|^{2}+V_{z}^{2}}$, and normalized eigen-vectors  $m=(m_{\uparrow},m_{\downarrow})$ for $\epsilon_{-}$ eigen-function and $p=(p_{\uparrow},p_{\downarrow})$ for $\epsilon_{+}$ eigen-function with:
\begin{eqnarray}\label{vectm}
m_{\uparrow}=\frac{-\tilde{t}_{so}}{[2R(R+V_{z})]^{1/2}}, &
m_{\downarrow}=\frac{R+V_{z}}{[2R(R+V_{z})]^{1/2}},\\\label{vectp}
p_{\uparrow}=\frac{-\tilde{t}_{so}}{[2R(R-V_{z})]^{1/2}}, &
p_{\downarrow}=\frac{-(R-V_{z})}{[2R(R-V_{z})]^{1/2}}.
\end{eqnarray}
Next we introduce chiral basis; with the help of Eqs.~(\ref{vectm}) and (\ref{vectp}) for spin $\sigma=\uparrow,\downarrow$ fermionic operator:
\begin{equation}
c_{k\sigma}=m_{\sigma}c_{k-}+p_{\sigma}c_{k+},
\end{equation}
where fermionic operators $c_{k\mp}$ annihilate states with momentum $k$ in the lower/upper chiral band. The Hamiltonian Eq.~(\ref{H_in_k}) written in the chiral basis assumes the form $H=H_{-}+H_{+}+H_{+-}$, where:
\begin{eqnarray}\label{H_mp}\nonumber
H_{\mp}=\sum_{k}\epsilon_{k\mp}c^{\dagger}_{k\mp}c_{k\mp}+\\
\sum_{k}[\Delta_{\mp}c_{k\mp}c_{-k\mp}+\Delta^{\star}_{\mp}c_{-k\mp}^{\dagger}c_{k\mp}^{\dagger}],
\end{eqnarray}
$\Delta(k)_{\mp}=\pm\tilde{t}_{so}(k)\Delta/(2R(k))$, and
\begin{equation}\label{H+-}
  H_{+-}=\sum_{k}[\Delta(k)_{+-}c_{k+}c_{-k-}-\Delta(k)_{+-}^{\star}c_{-k-}^{\dagger}c_{k+}^{\dagger}],
\end{equation}
with $\Delta(k)_{+-}=i\Delta V_{z}/R(k)$. We have indicated here explicitly the $k$-dependence of superconducting order parameters.

For the following input parameters in the wire \cite{Stefanski2021}; the induced $s$-wave order parameter $\Delta=0.2$, chemical potential $\mu=0$, Rashba interaction $t_{so}=2\Delta$, magnetic field $V_{z}=1.2 V_{z}^{cr}=1.2\Delta$, and under approximation $\overline{\sin^{2}(ka)}=1/2$ (lattice constant $a\equiv 1$), we obtain an intra-band order parameter $|\Delta_{\mp}|\simeq 0.86 \Delta$ and an inter-band order parameter $|\Delta_{+-}|\simeq\Delta/3$. It is worth emphasizing that in order to spot the MZM from the upper chiral sub-band the QD-wire coupling does not need to be of the order of the chiral sub-bands splitting. For the present parameters, the average splitting is $\epsilon_{+}-\epsilon_{-}\simeq 6\Delta$ whereas the QD level broadening is $\Gamma_{s}=\pi|t_{w}|^{2}\rho_{w}=0.75\Delta$ (with approximated density of states of a $1D$-wire $\rho_{w}=1/4$), confirming that the main information on MZMs is confined at Fermi energy.
\section{Toy Model}
In order to emphasize basic physics behind the general model, we introduce a toy model.
The description of the  bare dot by Eq.~(\ref{H_QD}) and its matrix representation $\hat{g}_{0}$ is retained as in the numerical model, whereas the topological wire is modelled by the simplest low energy Hamiltonian of two chiral sub-bands, $"-"$ and $"+"$, with Majorana zero modes formed in each sub-band. Generally the Majorana operator has the form of a $p$-wave Bogoliubov quasiparticle:
\begin{equation}\label{MZMquasi}
  \gamma_{i}=\sum_{\sigma=\uparrow,\downarrow}\int dx\left( u_{i,\sigma}(x)\Psi_{\sigma}(x)+v^{\star}_{i,\sigma}(x)\Psi^{\dagger}_{\sigma}(x)\right),
\end{equation}
where $\Psi_{\sigma}(x)$ is the electron field operator of the wire. Majorana operators have the property $\gamma_{i}=\gamma_{i}^{\dagger}$ ($i=1,2)$, which implies $u_{i,\sigma}(x)=v^{\star}_{i,\sigma}(x)$. The Majorana modes at the ends of the wire have two spin components: $\gamma_{i}=\gamma_{i,\downarrow}+\gamma_{i,\uparrow}$, forming Majorana Kramer's pair. When external magnetic field is applied, two chiral sub-bands are created, and each member of the pair  belongs to the sub-band of different chirality. The member of the pair from the lower sub-band (determined by direction of the magnetic field) becomes dominant and visible in tunneling spectroscopy. The second member of the MZM pair remains "dark" until spotted by QD-wire Andreev tunnelling of proper chirality.

The sub-band chirality, defined by the direction of the spin of electron relative to its $k$-wave vector direction,
in the $0D$-dimensional dot becomes spin direction with spin-chirality correspondence: $\downarrow\rightarrow "-"$ and $\uparrow\rightarrow "+"$.
Similarly, in the Hamiltonian, Eqs.~(\ref{H_mp})-(\ref{H+-}), for $k\rightarrow 0$ the same correspondence is retained. Thus, in the Toy Model, it is sufficient to retain only spin indices in the description, with understanding their general meaning in the wire.

The Hamiltonian of the wire is described by hybridized Majoranas in each chirality sector:
\begin{equation}\label{H_M}
  H_{M}=i\epsilon_{M}\sum_{j=\downarrow,\uparrow}\gamma_{1,j}\gamma_{2,j}.
\end{equation}

The dot is coupled to the adjacent MZMs in the wire in each chiral sector, described the Hamiltonian $H_{QD-M}$:
\begin{equation}\label{H_QD_M}
  H_{QD-M}=t_{M}^{-}(d^{\dagger}_{\downarrow}-d_{\downarrow})\gamma_{1,\downarrow}+t_{M}^{+}(d^{\dagger}_{\uparrow}-d_{\uparrow})\gamma_{1,\uparrow},
\end{equation}
where indices $"-"$ and $"+"$ denote the appropriate chiral sub-bands.

Our goal is to construct equation as Eq.~(\ref{GdMat}), with
the selfenergy matrix $\hat{\Sigma}_{wire}$ calculated within the low energy effective model and  $\hat{g}_{0}$ and $\hat{\Sigma}_{l}$ being the same as in general model.

We construct the matrix of the Green's functions of MZMs coupled to the dot, $\langle\langle\gamma_{1,\downarrow(\uparrow)}|\gamma_{1,\downarrow(\uparrow)}\rangle\rangle$, residing in the $"-"$ $("+")$ chiral sub-band. From the set of EOMs generated using Eqs.~(\ref{H_M}) and (\ref{H_QD_M}) we obtain the MZM $\hat{M}_{s}$ matrix. Expressed in the basis $\Psi=(\gamma_{1,\downarrow},\gamma_{1,\uparrow},\gamma_{1,\uparrow},\gamma_{1,\downarrow})$, it has the form:
\begin{eqnarray}
\hat{M}_{s}=\frac{\omega}{\omega^2-{\epsilon_{M}}^{2}} \left(
\begin{array}{cccc}
1 & 0 & 0 & 1\\
0 & 1 & 1 & 0\\
0 & 1 & 1 & 0\\
1 & 0 & 0 & 1
\end{array}
\right).
\end{eqnarray}
The QD-MZM coupling martix $\hat{t}_{M}$ is diagonal with the elements $\hat{t}_{M}[1,1]=-\hat{t}_{M}[4,4]=t_{M}^{-}$ and $\hat{t}_{M}[2,2]=-\hat{t}_{M}[3,3]=t_{M}^{+}$.

The selfenergy introduced by the coupling of the QD to the wire is then: $\hat{\Sigma}_{wire}=\hat{t}_{M}\hat{M}\hat{t}_{M}^{\star}$.
This yields the Dyson equation, Eq.~(\ref{GdMat}), for the Toy Model.
\begin{figure}
\centering
\includegraphics[width=9cm]{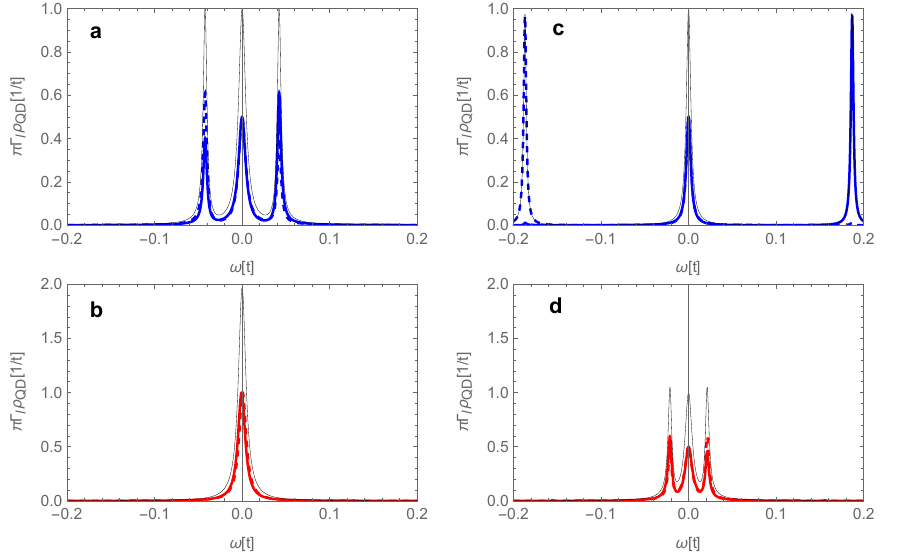}
\caption{\label{fig1toy}Spectral densities of the dot calculated within the Toy Model. Solid (dashed) line - particle (hole) parts, thin solid line - total spectral densities. Panels (a) and (c) are for spin-$\downarrow$ sector, when the dot spin level $\epsilon_{d\downarrow}=\epsilon_{F}$ for $\epsilon_{d}=0.123$, while panels (b) and (d) are for spin-$\uparrow$ sector, when the dot spin level $\epsilon_{d\uparrow}=\epsilon_{F}$ for $\epsilon_{d}=-0.123$. Panels (a) and (b) are for the dot weakly coupled to the wire for $t_{M}^{-}=0.03$ and $t_{M}^{+}=0$. Panels (c) and (d) are for the dot strongly coupled to the wire for $t_{M}^{-}=0.15$ and $t_{M}^{+}=0.015$. Remaining input parameters are: $\mu=0$, $V_{z}=1.2V_{z}^{cr}$, $V_{z}^{QD}=0.5 V_{z}$, $\epsilon_{M}=0$ and $\Gamma_{l}=0.005$.}
\end{figure}

In Fig.~(\ref{fig1toy}) the spectral densities of the dot, calculated within Toy Model, are represented.
The arrangement of the graphs directly corresponds to the one in Fig.~(\ref{fig1U=0}). For the weak coupling (panel (a)) a characteristic three peak structure is observed with MZM resonance in the middle and two satellite peaks at $\omega\sim\mp t_{M}^{-}$. At the same time, when $\epsilon_{d\uparrow}=\epsilon_{F}$ (panel (b)), a trivial resonance is only present at Fermi energy, because the upper chiral sub-band is decoupled from the dot.

In turn, for strong QD-wire coupling  (column (c)-(d)), the satellite peaks in spin down spectral density (panel (c)) are shifted far apart from MZM resonance, here we retained them the gap for illustration. Now at the same time the dot is also coupled to the $"+"$-chiral sub-band, which signals the appearance of the MZM resonance in the spin-up sector (panel (d)), with characteristic satellite peaks at $\omega\sim\mp t_{M}^{+}$  due to the splitting of the $\epsilon_{d\uparrow}$ level.

Contrary, for the case of the wire in its trivial state, only a single peak would be present at Fermi energy, when any of the QD spin-levels is tuned to $\epsilon_{F}$, regardless of the strength of QD-wire coupling.
\section{Conclusions}
In the present work we proposed a supplementary evidence, which complements the Majorana modes formation manifested by zero bias peak in differential conductance. Namely, in the setup of a quantum dot attached to the topological superconducting wire, we demonstrated that the formation of chiral sub-bands in the density of states of the wire can be utilized for Majorana modes detection. The presence of such sub-bands is an indication of the strong spin-orbit interaction in the wire heterostructure, a crucial ingredient for MZM to appear. Moreover, a quantum dot, unintentionally formed in the tunneling junction, which usually hinders MZM detection by introducing trivial Andreev levels, can be successfully utilized in conjunction with the chiral sub-bands for revealing MZMs. The characteristic three peak structures appearing when the dot spin sub-levels are tuned to Fermi energy, recognizable in either  sub-band chirality, indicate MZMs. Namely, for the dot strongly coupled to the wire, the satellite peaks of the dot sub-level  $\epsilon_{\sigma}$  split  by MZM can be moved outside the gap, and the remaining single MZM resonance is hardly distinguishable from trivial Andreev resonance. However additionally, the analogous resonances indicating MZM are still observable for $\epsilon_{\bar{\sigma}}=\epsilon_{F}$ in the other spin sector, weaker coupled to the second chiral sub-band, which considerably strengthens the attempt of Majorana mode detection.

It is interesting to note that in the case of the accessible Majorana zero modes belonging to the both chiral sub-bands, the possibility opens up to preform braiding operations on MZMs in each chirality. It considerably enriches Majorana-based quantum computing scheme \cite{Marra2022d}.

An important question arises how this picture changes by including Coulomb interactions in the dot. The singlet-doublet quantum phase transition in the interacting dot coupled to $s$-wave superconductor  \cite{Valentini2021}, will be significantly altered in case of $p$-wave superconducting wire due to the equal-spin superconducting pairs entering the dot. We also expect a Coulomb mediated interaction between chiral sub-band, which may influence the three-peak structures appearing in the spectral density of the dot. These problems are addressed in the follow-up work.

\bibliography{PStefPM23arXiv}

\end{document}